\documentclass[aps,twocolumn,preprintnumbers,nofootinbib,superscriptaddress]{revtex4-1}
\usepackage{amsmath} \usepackage{graphicx} \usepackage{amsfonts}
\usepackage{array} \usepackage{amsthm} \usepackage{bm}
\usepackage{palatino} \usepackage{mathpazo} %\usepackage{mathptmx}
\usepackage{supertabular}

\usepackage[breaklinks]{hyperref}
\usepackage{color}

\newcommand{\be}{\begin{equation}}
\newcommand{\ee}{\end{equation}}
\newcommand{\ba}{\begin{eqnarray}}
\newcommand{\ea}{\end{eqnarray}}
\newcommand{\bal}{\begin{align}}
\newcommand{\eal}{\end{align}}

\newcommand{\dd}{\text{d}}

\newcommand{\bb}{\bibitem}

\newcommand{\bw}{\begin{widetext}}
\newcommand{\ew}{\end{widetext}}

\begin{document}
%\begin{flushleft}???
%\end{flushleft}
%\title{Logarithmically singular NED black holes with(out) de Sitter term}
\title{Two-sided NED black-hole, naked-singularity, and soliton solutions}

\author{Mustapha Azreg-A\"{\i}nou}%\email{?????}
\affiliation{Ba\c{s}kent University, Engineering Faculty, Ba\u{g}l{\i}ca Campus, Ankara, Turkey}

%\date{}

\begin{abstract}
We consider non-linear electrodynamics (NED) minimally coupled to general relativity. We derive novel electrically charged, spherically symmetric, black-hole solutions having, for some set of parameters, all their NED fields (the electric field and the square of the electromagnetic field) regular for all values of the radial coordinate. For another set of parameters, the NED fields and the Kretschmann scalar are regular as the radial coordinate runs from one spatial infinity to another spatial infinity without the metric being a wormhole. We obtain solutions that have two distinct or the same asymptotic behaviors (two spatial infinities) with equal or unequal ADM masses and solutions with always one horizon whatever the ratio of the electric charge to mass. We comment on some regularity theorems and generalize them to multi-valued NED Lagrangians. The derived regular solutions do not violate the Weak Energy Condition.
\end{abstract}

%\pacs{}

\maketitle

\section{Introduction}
Regular and non-regular black holes (BH) in general relativity (GR) and its extensions coupled to non-linear electrodynamics (NED)  have attracted, and continue to attract, much attention in the scientific community~\cite{Dymni15,Odi17,Gao21,2211.00743,2405.09124,2412.00582,2409.14946,2409.07305,2403.02056,Faizuddin} and references therein. The regularity conditions of the NED solutions are solely imposed on the scalar invariants of the metric and not on the NED fields~\cite{2211.00743}. 

Among the theorems on their regularity, \cite{2211.00743} and references therein, one states that if the NED Lagrangian $\mathcal{L}(F)$ approaches the Maxwellian value $F$, in the limit of small $F$, no center-regular static, spherically symmetric, electrically charged metric solution exits. Nothing is said about the regularity of the NED fields (the electric field and the square of the electromagnetic field tensor): The theorem does not exclude solutions with regular NED fields at the center and elsewhere and that is what we are going to show in this letter. Our main aim is to derive novel solutions where both the metric and the NED fields are not only regular at the center, but everywhere; that is, fully regular solutions. Moreover, the theorem does not consider multi-valued NED Lagragians $\mathcal{L}(F)$. By spherical symmetry, a magnetic-field solution has its NED fields diverging at the center.

In Sec.~\ref{secfe} we introduce the field equations and the working ansatz. In the subsequent Secs.~\ref{secn2}-\ref{secn3} we derive a variety of NED solutions along with the spacetime metrics upon specializing to physically interesting, and mathematically tractable, cases. We will obtain novel solutions that have two different or the same asymptotic behaviors with equal or unequal ADM masses as well as fully regular solutions. In Sec.~\ref{secwek} we will show that the derived regular solutions do not violate the Weak Energy Condition and we will conclude in Sec.~\ref{seccon}.

\section{Fields equations\label{secfe}}
Let $F_{\mu\nu}$ be the electromagnetic field tensor and $F=F_{\mu\nu}F^{\mu\nu}$. If $\mathcal{L}(F)$ represents the NED Lagragian, the filed equation are derived by varying the action
\begin{equation}\label{1}
S=\frac{1}{2}\int \dd ^4x\sqrt{|g|}~[R-\mathcal{L}(F)]\,,
\end{equation}
where we took $c= 8\pi G=1$. If we choose the static, spherically symmetric, metric of the form
\begin{equation}\label{2}
	\dd s^2=f(r)\dd t^2 - \frac{1}{f(r)}\dd r^2 - r^2\dd \Omega^2\,,
\end{equation}
then the field equations spell as
\begin{align}
\label{3a}&\frac{rf'+f-1}{r^2}=-\frac{\mathcal{L}}{2}+F\mathcal{L}_F\,,\\
\label{3b}&\frac{rf''+2f'}{2r}=-\frac{\mathcal{L}}{2}\,,\\
\label{3c}&\nabla_{\mu}(\mathcal{L}_F~F^{\mu\nu})=0\,,
\end{align}
where $\mathcal{L}_F=\dd \mathcal{L}/\dd F$ and the prime notation denotes derivative with respect to $r$. The Kretschmann scalar $KS=R_{\mu\nu\rho\sigma}R^{\mu\nu\rho\sigma}$ related to the metric~\eqref{2} has the form
\begin{equation}\label{KS}
	KS=\frac{4 [1-f(r)]^2+4 r^2 \left(f(r)'\right)^2+r^4 \left(f(r)''\right)^2}{r^4}\,.
\end{equation}
For static, spherically symmetric metrics, it was shown in~\cite{2211.00743}, that the KS is a quadratic positive sum of four terms, of the form $\sum_{i=1}p_iK_i^2$ and $p_i>0$, and all other algebraic scalar curvature invariants are linear, cubic, and, in general, integer-power combinations of the same $K_i$'s. Thus, if the KS is regular at some point (or everywhere), all the $K_i$'s are regular, and consequently, all other algebraic scalar curvature invariants are regular too.

Looking for an electrically charged solution~\eqref{2}, Eq.~\eqref{3c} yields the electric field expression along with that of the invariant $F$
\begin{equation}\label{4}
F^{tr}=	\frac{q}{r^2\mathcal{L}_F}\,,\quad F=-2(F^{tr})^2=-\frac{2q^2}{r^4\mathcal{L}_F^2}\,. 
\end{equation}

There are many ways to remove the singularity of the electric field~\eqref{4} at the center $r=0$ taking, for instance, $\mathcal{L}_F= \coth (r/K)$, or preferably, of the form
\begin{equation}\label{5}
	\mathcal{L}_F=\frac{K^n}{r^n}+1=\frac{1}{x^n}+1\,,\quad n\geq 2\,,
\end{equation}
where $K$ is a positive parameter having the dimension of length, which will be given in term of the ratio $q^2/M$, with $M$ being the ADM mass of the solution and $x=r/K$ is a dimensionless radial coordinate.
%Among the cases that are analytically tractable, we will consider the case $n=4$; the case $n=2$ yields the trivial flat solution.

Equation~\eqref{5} is not in its most general form and---as we shall see---the inclusion of other powers, 
\begin{equation}\label{5p}
	\mathcal{L}_F=\frac{1}{x^n}+\sum_{i=1}^{n-1}\frac{h_{n-i}}{x^{n-i}}+1\,,\quad n\geq 2\,,
\end{equation}
where $h_{n-i}$ are dimensionless real parameters, may simplify the final expression of the metric solution.

Note that the independent term in~\eqref{5}, and in~\eqref{5p}, has been taken to be $1$ so that, by integration, $\mathcal{L}$ will have a Maxwellian behavior in the limit $F\to 0$, which corresponds to the limit $x\to\infty$ (the weak-field limit or the linear limit):
\begin{equation}\label{5pp}
	\mathcal{L}(F)=\text{const}+F+\cdots\,,\quad x\to\infty\,.
\end{equation}
In the vicinity of the center $x\to 0$ (the non-linear limit), $\mathcal{L}$ need not be Maxwellian. Note that if $n>2$, $F\to 0$ in both limits $x\to 0$ and $x\to\infty$. In this case we obtain in the limit $x\to 0$ %By integration of~\eqref{4} and~\eqref{5} we obtain for $n>2$
%\begin{multline}
%	\mathcal{L}=\text{const}-\frac{2n-4}{n-4}~\Big(\frac{K^4}{2q^2}\Big)^{-n/(2n-4)}~\big(-F\big)^{(n-2)/(2n-4)}\\
%	+\cdots\,,\quad x\to 0\,,\quad (n>2)\,.
%\end{multline}
\begin{equation}\label{5ppp}
	\mathcal{L}(F)\simeq\begin{cases}
		\text{const}-\frac{2n-4}{n-4}\Big(\frac{2q^2}{K^4}\Big)^{\frac{n}{2n-4}}\big|F\big|^{\frac{n-2}{2n-4}}, & n>2\,,n\neq 4\,,\\
		\text{const}-\Big(\frac{2q^2}{K^4}\Big)\ln|F|, & n= 4\,.
	\end{cases}
\end{equation}
In both Eqs.~\eqref{5pp}-\eqref{5ppp}, $F\to 0$. We are thus dealing with a multi-valued Lagragian $\mathcal{L}$ seen as a function of $F$. For $n=2$, the expansion of $\mathcal{L}(F)$ (also a multi-valued function of $F$) in the limit $F\to 0$, is given in the text.\vskip4pt
%In the case $n=2$, $\mathcal{L}\to (F+2q^2/K^4)^{-1}$ in the limit $x\to 0$ if $h_1\neq 0$ and  $\mathcal{L}\to \ln|F+2q^2/K^4|$ in the limit $x\to 0$ if $h_1= 0$.

On applying Gauss's Theorem for a sphere of radius $r$ and center at $r=0$, we see from~\eqref{4} and~\eqref{5p} that the flux of the electric field yields the value $q/\epsilon_0$ only in the limit of large $r$. This means that the charge $q$ is not local, rather distributed on a space region that extends to spatial infinity.

\section{The case $n=2$\label{secn2}}
Let $\mathcal{L}_F$ be of the form~\eqref{5p}
\begin{equation}\label{1n2}
	\mathcal{L}_F=\frac{1}{x^2}+\frac{h}{x}+1\,,
\end{equation}
where $h$ is a dimensionless parameter such that $1+hx+x^2=0$ has no real roots or has no positive roots so that the electric field $F^{tr}$ and the invariant $F$ are finite for all $x\geq 0$. We obtain
\begin{equation}\label{2n2}
	F^{tr}=	\frac{q}{K^2(1+hx+x^2)}\,,\quad F=\frac{-2q^2}{K^4(1+hx+x^2)^2}\,, 
\end{equation}
{\small
\begin{align}
&\mathcal{L}=\frac{4 q^2}{K^4} \Big[\frac{-h+(1-2 h^2) x-2 h x^2}{x (1+h x+x^2)}-(h^2-1) \ln \Big(\frac{x^2}{1+h x+x^2}\Big)\nonumber\\
\label{3n2a}&+\frac{2 h (h^2-3)}{\sqrt{4-h^2}}\arctan \Big(\frac{h+2
	x}{\sqrt{4-h^2}}\Big)+c\Big]\,
\end{align}}%
where $c$ is a dimensionless constant of integration. The case $h= 2$ along with the case $h=-2$ (where $1+hx+x^2=0$ has a double positive root) will be treated separately. One can express $x$ in terms of $F$ upon reversing the second equation in~\eqref{2n2}, then one substitutes into~\eqref{3n2a}, the obtained expression of $\mathcal{L}$ has the following series expansion in the Maxwellian limit $F\to 0$
\begin{multline}\label{3n2b}
\mathcal{L}(F)=	\frac{4 [c \sqrt{4-h^2}-\eta h (3-h^2) \pi ] q^2}{\sqrt{4-h^2} K^4}+F\\-\frac{\eta 2^{7/4} h K |F|^{5/4}}{5 \sqrt{|q|}}+\mathcal{O}(|F|^{3/2})\,,\quad\eta =\pm 1\,,
\end{multline}
where $\eta =+1$ corresponds to $x\to +\infty$ and $\eta =-1$ corresponds to $x\to -\infty$ [in this case too $\mathcal{L}(F)$ is a multi-valued function of $F$]. The first term represents twice a cosmological constant and it vanishes if
\begin{equation}\label{cos}
	c=\frac{\eta h (3-h^2) \pi}{\sqrt{4-h^2}}\,,\quad h\neq \pm 2\,.
\end{equation}

We see from the expression of $\mathcal{L}$ that the cases $h=\epsilon$, $h=\epsilon \sqrt{3}$ ($\epsilon =\pm 1$), and $h=0$ offer important simplifications. The general metric function $f(r)$ in terms of $|h|< 2$ is given in the Appendix by~\eqref{A0} and the case $h= \pm 2$ is treated separately. In this section we focus on these three special and simple cases.\vskip4pt

\begin{figure}[!htb]
	\centering
	\includegraphics[width=0.75\linewidth]{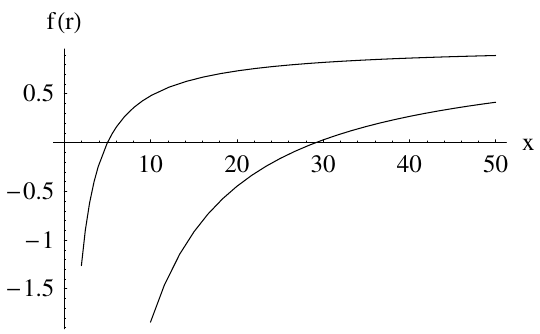}
	\caption{Novel BH: One horizon for any value of $q^2/M^2$. Plots of $f(r)$~\eqref{3n2} corresponding to the flat metric with $\epsilon =+1$ ($c=2\pi/\sqrt{3}$). Upper plot: $4\pi^2q^2/(27M^2)=4/9$. Lower plot: $4\pi^2q^2/(27M^2)=4/49$.}
	\label{Fign2hp1}
\end{figure}

\begin{figure}[!htb]
	\centering
	\includegraphics[width=0.75\linewidth]{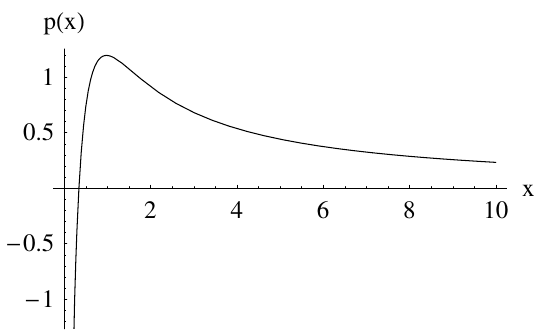}\\
	\includegraphics[width=0.75\linewidth]{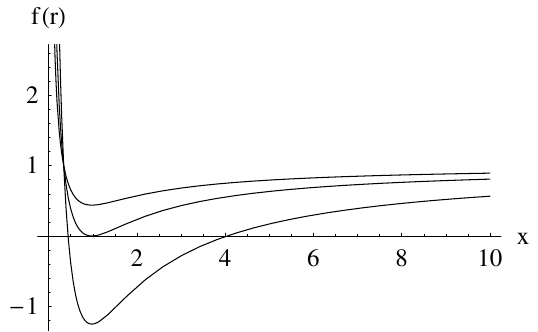}
	\caption{Plots of $p(x)$ and $f(r)$~\eqref{3n2} corresponding to the flat metric with $\epsilon =-1$ ($c=-2\pi/\sqrt{3}$). In the lower plot (from up to down) we have a NS ($4\pi^2q^2/(27M^2)>1.19415$), a BH with one horizon ($4\pi^2q^2/(27M^2)=1.19415$), and a BH with two horizons ($4\pi^2q^2/(27M^2)<1.19415$).}
	\label{Fign2hm1}
\end{figure}

For $h=\epsilon$, we have $F\mathcal{L}_F=-2q^2/[K^4x^2(1+\epsilon x+x^2)]$, along with~\eqref{3n2a}, Eqs.~\eqref{3a} and ~\eqref{3b} are solved by
\begin{align}\label{3n2}
f(x)=&1-\frac{q^2}{K^2}~p(x)=1-\frac{27M^2}{4\pi^2 q^2}~p(x)\,,\nonumber\\
f(r)=&1-\frac{2 q^2}{3 K^2} \Big[(1-2 \epsilon  x+c x^2)\nonumber\\
\quad &+\frac{4 (1-\epsilon  x^3)}{\sqrt{3} x} \arctan \Big(\frac{2 x+\epsilon }{\sqrt{3}}\Big)\Big]\,.	
\end{align}
This metric is singular at the center $x=0$
{\small
\begin{equation}
	f(r)=\frac{-8 q^2 \arctan (\epsilon /\sqrt{3})}{3 \sqrt{3} K^2 x}+\Big(1-\frac{2 q^2}{K^2}\Big)+\frac{2 q^2 \epsilon  x}{K^2}+\mathcal{O}(x^3)\,,
\end{equation}}%
and its Kretschmann scalar diverges as $KS=64\pi^2 q^4/(81K^8x^6)+\mathcal{O}(x^{-5})$. The metric behaves asymptotically as
\begin{multline}\label{h1p}
f(r)=	\frac{2 q^2 (2 \sqrt{3} \pi  \epsilon -3 c) x^2}{9 K^2}+1-\frac{4 \pi  q^2}{3 \sqrt{3} K^2 x}\\+\frac{q^2}{K^2 x^2}+\mathcal{O}(x^{-3})\,,\quad x\to +\infty\,,
\end{multline}
which is flat if $c=2  \pi  \epsilon / \sqrt{3}$~\eqref{cos}, de Sitter if $c>2 \pi  \epsilon / \sqrt{3}$, and anit-de Sitter if $c<2 \pi  \epsilon / \sqrt{3}$. From the Schwarzschild term in~\eqref{h1p}, that is, the coefficient of $1/r$, we obtain the ADM mass $M$ satisfying $2M=4\pi q^2/(3\sqrt{3}K)$, which can be reversed to give $K=2\pi q^2/(3\sqrt{3}M)$. For $c=2  \pi  \epsilon / \sqrt{3}$ and $\epsilon =1$ the function $p(x)$~\eqref{3n2} is monotonically decreasing and it has no upper limit as it diverges in the limit $x\to 0$, so the metric~\eqref{3n2} represents a BH \emph{with one and only one} horizon for any value of the electric charge (precisely for any value of $q^2/M^2$). This represents a novel BH as this is not met in Maxwellian electrodynamics. For $\epsilon =-1$ the function $p(x)$~\eqref{3n2} has a maximum value of $1.19415$ at $x=0.981739$, so the metric~\eqref{3n2} represents a horizonless solution (naked singularity: NS) if  $K^2/q^2=4\pi^2q^2/(27M^2)>1.19415$ [in this case the horizontal line $y=K^2/q^2$ does not intersect the graph of $y=p(x)$], an extreme BH with one horizon if $4\pi^2q^2/(27M^2)=1.19415$ [in this case the line $y=K^2/q^2$ intersect the graph of $y=p(x)$ at one point], and a BH with two horizons if $4\pi^2q^2/(27M^2)<1.19415$ [in this case the line $y=K^2/q^2$ intersects the graph of $y=p(x)$ at two points]. Illustrations are shown in Figs.~\ref{Fign2hp1} and ~\ref{Fign2hm1}.
\begin{figure*}[!htb]
	\centering
	\includegraphics[width=0.43\linewidth]{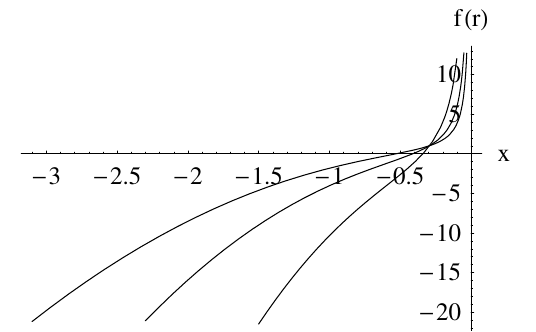}
	\includegraphics[width=0.43\linewidth]{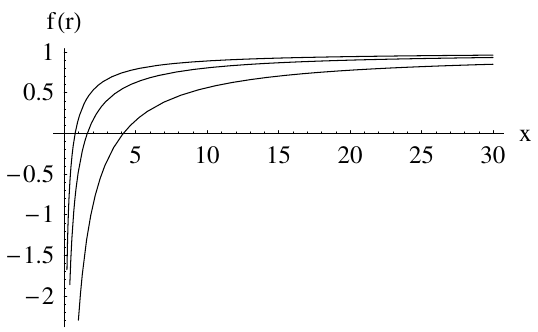}
	\caption{$n=2$ and $h=1$. LS (left-side) and RS (right-side) BHs. Plots of $f(r)$~\eqref{3n2} corresponding to $\epsilon =+1$ and $c=2\pi/\sqrt{3}$ using the same values of $q^2/M^2$ for the left and right plots. Right plot: Represents a flat BH~\eqref{h1p} with only one horizon (this is a reproduction of Fig.~\ref{Fign2hp1} for different values of $q^2/M^2$). Left plot: Represents an asymptotically de Sitter metric~\eqref{h1m}.}
	\label{Figrbhfh1}
\end{figure*}
\begin{figure*}[!htb]
	\centering
	\includegraphics[width=0.43\linewidth]{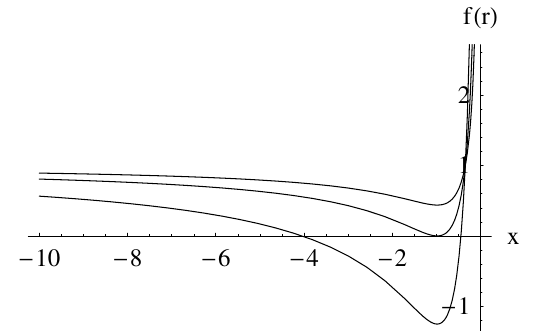}
	\includegraphics[width=0.43\linewidth]{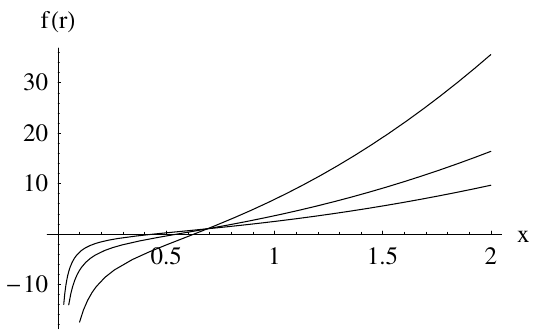}
	\caption{$n=2$ and $h=1$. LS (left-side) solution and RS (right-side) BH. Plots of $f(r)$~\eqref{3n2} corresponding to $\epsilon =+1$ and $c=-2\pi/\sqrt{3}$ using the same values of $q^2/M^2$ for the left and right plots. Right plot: Represents an asymptotically anti-de Sitter metric~\eqref{h1m}. Left plot: Represents a flat solution~\eqref{h1p} with no horizon, with one double horizon (extreme BH), and with two horizons for different values of $q^2/M^2$.}
	\label{Figlbhfh1}
\end{figure*}
\begin{figure}[!htb]
	\centering
	\includegraphics[width=0.99\linewidth]{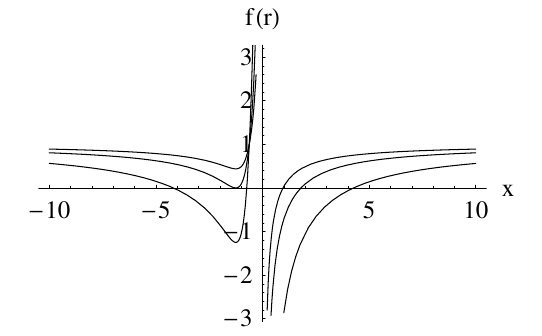}
	\caption{$n=2$ and $h=\epsilon\sqrt{3}$. LS (left-side) solution and RS (right-side) BH. Plots of $f(r)$~\eqref{3n22} corresponding to $\epsilon =+1$ and $c=0$ using the same values of $q^2/M^2$ for the left and right plots. Both LS solution~\eqref{h3m} and RH BH~\eqref{h3p} are asymptotically flat having the same mass.}
	\label{Figlrbhfh3}
\end{figure}

The metric function~\eqref{3n2} is valid for $x<0$ too as it solves the field Eqs.~\eqref{3a}-\eqref{3b}, there is thus no reason to restrict the radial coordinate $r$ to positive values only and this yields the maximum extent of the coordinate chart for the manifold: $r\in (-\infty,0)\cup (0,\infty)$, excluding $r=0$ where $f$ and $KS$ are singular. Extending $r$ to negative values and beyond singularities~\cite{HM} is a new fact in NED. The metric function~\eqref{3n2} behaves asymptotically in the limit $x\to -\infty$ as
\begin{multline}\label{h1m}
f(r)=-	\frac{2 q^2 (2 \sqrt{3} \pi  \epsilon +3 c) x^2}{9 K^2}+1-\frac{4 \pi  q^2}{3 \sqrt{3} K^2 (-x)}\\+\frac{q^2}{K^2 x^2}+\mathcal{O}(x^{-3})\,,\quad x\to -\infty\,.
\end{multline}
This is another novel behavior, not observed in the Maxwellian limit of the NED theory. An instance of that is the Reissner-Nordstr\"om and the Schwarzschild metric which each behaves in the limit $r\to -\infty$ like a naked singularity with a negative mass $-M$. The metric function~\eqref{3n2} behaves asymptotically in the limit $x\to -\infty$ still as a charged BH with positive mass~\eqref{h1m} that is given by the same expression $M=2\pi q^2/(3\sqrt{3}K)$ as that corresponding to the limit $x\to +\infty$ .

In order to illustrate graphically the two-side solution, we focus on the case $h=1$ ($\epsilon=+1$). Note that when right-side (RS) solution is flat ($c=2\pi/\sqrt{3}$)~\eqref{h1p}, the left-side (LS) solution has a de Sitter asymptotic behavior~\eqref{h1m}. This is clearly illustrated in Fig.~\ref{Figrbhfh1} where the right plot is a reproduction of Fig.~\ref{Fign2hp1} using different values of $4\pi^2q^2/(27M^2)$. In Fig.~\ref{Figlbhfh1} we take $c=-2\pi/\sqrt{3}$, so that the LS BH is asymptotically flat~\eqref{h1m} and the RS BH has an anti-de Sitter asymptotic behavior~\eqref{h1p}. As we shall see in the subsequent case and sections, we will obtain LS and RS solutions with the same behavior (both asymptotically flat) for the same values of the parameters.

For $h=\epsilon$, the NED fields are regular for all $-\infty<x<\infty$.\vskip4pt

For $h=\epsilon\sqrt{3}$, the NED fields are also regular for all $-\infty<x<\infty$. We have $F\mathcal{L}_F=-2q^2/[K^4x^2(1+\epsilon\sqrt{3} x+x^2)]$, along with~\eqref{3n2a}, Eqs.~\eqref{3a} and ~\eqref{3b} are solved by
\begin{align}\label{3n22}
	f(x)=&1-\frac{q^2}{K^2}~p(x)=1-\frac{9M^2}{4\pi^2 q^2}~p(x)\,,\nonumber\\
	f(r)=&1-\frac{q^2}{K^2}\Big[\frac{2}{3}+\frac{2 c x^2}{3}-\frac{4 x \epsilon }{\sqrt{3}}\\
	\quad &+\frac{8 \arctan (2 x+\sqrt{3} \epsilon )}{3
		x}+\frac{4}{3} x^2 \ln \Big(\frac{1+x^2+\sqrt{3} x \epsilon }{x^2}\Big)\Big]\,.\nonumber	
\end{align}
This metric is singular at the center $x=0$
\begin{multline}
		f(r)=-\frac{8 q^2 \pi  \epsilon }{9 K^2 x}+\left(1-\frac{2 q^2}{K^2}\right)+\frac{2 \sqrt{3} q^2 \epsilon  x}{K^2}\\
		-\frac{2 q^2 (4+3 c-12 \ln  |x|) x^2}{9
			K^2}+\mathcal{O}(x^3)\,,
\end{multline}
and its Kretschmann scalar diverges as $KS=256 \pi ^2 q^4/(27 K^8 x^6)+\mathcal{O}(x^{-5})$. The metric behaves asymptotically as
\begin{multline}\label{h3p}
		f(r)=	-\frac{2 c q^2 x^2}{3 K^2}+1-\frac{4 \pi  q^2}{3 K^2 x}+\frac{q^2}{K^2 x^2}\\
		+\mathcal{O}(x^{-3})\,,\quad x\to +\infty\,,
\end{multline}
which is flat if $c=0$~\eqref{cos}, de Sitter if $c>0$, and anit-de Sitter if $c<0$, with electric charge $q$ and ADM mass $M$ such that $K=2\pi q^2/(3M)$. For $c=0$ and $\epsilon =1$, the function $p(x)$~\eqref{3n2} is monotonically decreasing and it has no upper limit as it diverges in the limit $x\to 0$, so the metric~\eqref{3n22} represents a BH \emph{with one and only one} horizon for any value of the ratio $q^2/M^2$. For $c=0$ and $\epsilon =-1$ the function $p(x)$~\eqref{3n22} has a maximum value of $2.09842$ at $x=1.25118$, so the metric~\eqref{3n2} represents a NS if  $K^2/q^2=4\pi^2q^2/(9M^2)>2.09842$, an extreme BH with one horizon if $4\pi^2q^2/(9M^2)=2.09842$, and a BH with two horizons if $4\pi^2q^2/(9M^2)<2.09842$. The plots are similar to those of Figs.~\ref{Fign2hp1} and~\ref{Fign2hm1}. 

The metric function~\eqref{3n22} solves the field Eqs.~\eqref{3a}-\eqref{3b} for $x<0$ too and has an expansion as $x\to -\infty$, similar to Eq.~\eqref{h3p}, with the correct sign in front of $1/(-x)$, $-4 \pi  q^2/[3 K^2 (-x)]$,
\begin{multline}\label{h3m}
	f(r)=	-\frac{2 c q^2 x^2}{3 K^2}+1-\frac{4 \pi  q^2}{3 K^2 (-x)}+\frac{q^2}{K^2 x^2}\\
	+\mathcal{O}(x^{-3})\,,\quad x\to -\infty\,,
\end{multline}
yielding a LS solution for the same parameters. Illustrations of the LS and RS Bhs are shown in Fig.~\ref{Figlrbhfh3} taking $\epsilon =+1$ and $c=0$. Both the LS and RS solutions, having the same mass $M$, are asymptotically flat. 

Had we taken  $\epsilon =-1$ and $c=0$, we would have obtained the symmetric with respect to the $y$-axis of the plots in Fig.~\ref{Figlrbhfh3}. \vskip4pt

\begin{figure}[!htb]
	\centering
	\includegraphics[width=0.75\linewidth]{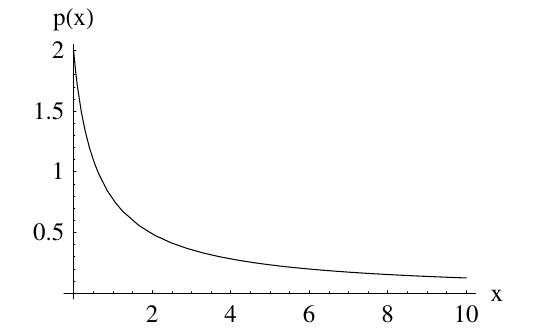}\\
	\includegraphics[width=0.75\linewidth]{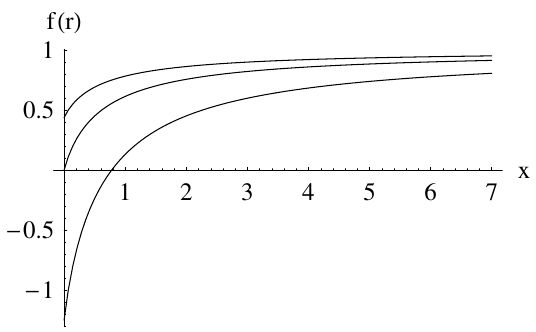}
	\caption{$n=2$ and $h=2$. Plots of $p(x)$ and $f(r)$~\eqref{3n23} corresponding to the flat metric with ($c=0$). The function $p(x)$~\eqref{3n23} has a finite limit $x\to 0$, $\lim_{x\to 0}p(x)=2$, and it is monotonically decreasing.. In the lower plot (from up to down) we have a (soft) NS ($4q^2/(9M^2)>2$), an extreme case where the horizon coincides with the center ($4q^2/(9M^2)=2$), and a BH with one horizon for all  $4q^2/(9M^2)<2$. In all cases the metric $f(r)$ has a finite value at the center given by $1-2q^2/K^2=1-9M^2/(2q^2)$.}
	\label{Fign2h2}
\end{figure}
\begin{figure}[!htb]
	\centering
	\includegraphics[width=0.99\linewidth]{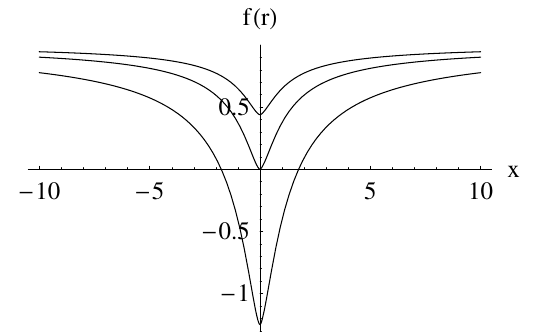}
	\caption{$n=2$ and $h=0$. LS (left-side) and RS (right-side) solutions. Plots of $f(r)$~\eqref{heq0} taking $c=0$ and using the same values of $q^2/M^2$ for the left and right plots. $f(r)$ is regular at the center and assumes a finite value there. Both LS solution and RH solution are asymptotically flat~\eqref{heq03} having the same mass. The plot shows soft LS and RS NSs, an extreme case, and LS and RS BHs each with one horizon only.}
	\label{Figlrbhfh0}
\end{figure}

As we shall see, the case $h=0$ has both LS and RS BHs, which become asymptotically flat for the same values of the parameters (as was the case $h=\epsilon\sqrt{3}$). The general metric function $f(r)$ along with its end-behaviors are given by 
\begin{align}\label{heq0}
&f(r)=1-\frac{q^2}{K^2}~p(x)=1-\frac{9M^2}{\pi^2 q^2}~p(x)\,,\\
&f(r)=1-\frac{2q^2}{3K^2} \Big[1+c x^2+\frac{2 \arctan  x}{ x}- x^2 \ln \Big(1+\frac{1}{x^2}\Big)\Big]\,.\nonumber
\end{align}
\begin{align}
	f=&\Big(1-\frac{2 q^2}{K^2}\Big)-\frac{2 q^2 (3 c-2+3 \ln  x^2) x^2}{9 K^2}+\frac{2 q^2 x^4}{5 K^2}\nonumber\\
\label{heq02}	\quad &+O(x^5)\,,\quad x\to 0\,,\\
f=&-\frac{2 c q^2 x^2}{3 K^2}+1-\frac{2 \pi  q^2}{3 K^2  |x|}+\frac{q^2}{K^2 x^2}\nonumber\\
\label{heq03}\quad &+O(x^{-3})\,,\quad x\to \pm \infty\,.
\end{align}
We see from~\eqref{heq03} that the metric function becomes flat in both directions $\pm\infty$ if we take $c=0$. The Kretschmann scalar diverges as $x^{-4}$ in the limit $x\to 0$. The function $p(x)$~\eqref{heq0} has a finite limit of $2$ as $x\to 0$ as is the function $f(r)$\eqref{heq02}. The plots of $p(x)$ and $f(r)$ for $x>0$ are the same as those shown in Fig.~\ref{Fign2h2} and the solution has a mass $M$ such that $K=\pi q^2/(3M)$. The solution may have only one horizon or none. As we see from~\eqref{heq03}, the case $h=0$ has both the LS and RS solutions (NSs or BHs) which are both flat if we choose $c=0$, as illustrated in Fig.~\ref{Figlrbhfh0}. 

The NED fields are regular for all $-\infty<x<\infty$.\vskip4pt

For $h= 2$, we have
\begin{equation}\label{n2h21}
\mathcal{L}=\frac{8 q^2}{K^4} \Big[-\frac{2+9 x+6 x^2}{2 x (1+x)^2}+3 \ln \Big|\frac{1+x}{x}\Big|+c\Big]\,,
\end{equation}
and $F\mathcal{L}_F=-2q^2/[K^4(x+x^2)^2]$, and Eqs.~\eqref{3a} and ~\eqref{3b} are solved by
\begin{align}\label{3n23}
f(x)=&1-\frac{q^2}{K^2}~p(x)=1-\frac{9M^2}{4q^2}~p(x)\,,\\
f(r)=&1-\frac{2 q^2}{3 K^2}\Big[3-6 x+2 c x^2+6 x^2 \ln \Big|\frac{1+x}{x}\Big|\Big]\,.\nonumber	
\end{align}
This metric is regular at the center $x=0$
{\small
	\begin{multline}\label{exp1}
	f(r)=\Big(1-\frac{2 q^2}{K^2}\Big)+\frac{4 q^2 x}{K^2}-\frac{4 q^2 (c-3 \ln  |x|) x^2}{3 K^2}+O\Big(x^3\Big)\,,
	\end{multline}}%
but its Kretschmann scalar diverges as $KS=16 q^4/(K^8 x^4)+\mathcal{O}(x^{-3})$. We see that in this case the singularity at the center has been softened, compared to the three previous cases ($h=\epsilon$, $h=\epsilon\sqrt{3}$, and $h=0$). The metric behaves asymptotically as
{\small
	\begin{equation}\label{exp2}
	f(r)=-\frac{4 c q^2x^2}{3 K^2}+1-\frac{4 q^2}{3 K^2 x}+\frac{q^2}{K^2x^2}+O(x^{-3})\,,
	\end{equation}}%
which is flat if $c=0$, de Sitter if $c>0$, and anit-de Sitter if $c<0$, with electric charge $q$ and ADM mass $M$ such that $K=2q^2/(3M)$. In the flat case ($c=0$), the function $p(x)$~\eqref{3n23} has a finite limit $x\to 0$, $\lim_{x\to 0}p(x)=2$, and it is monotonically decreasing. For $K^2/q^2=4q^2/(9M^2)>2$ the metric~\eqref{3n23} represents a (soft) NS and for $4q^2/(9M^2)<2$ the metric is a BH with one and only one horizon. The case with $4q^2/(9M^2)=2$ is an extreme case where the horizon coincides with the center at $x=0$. For all $c$, the metric function $f(r)$ assumes a finite value at the center given by $1-2q^2/K^2=1-9M^2/(2q^2)$~\eqref{exp1}. Illustrations are shown in Fig.~\ref{Fign2h2} for the flat case.

Substituting $x=-1+2^{1/4}\sqrt{|q|}/(K |F|^{1/4})$ into~\eqref{n2h21} we obtain $\mathcal{L}(F)$, which expands in the limit $F\to 0$ as
\begin{multline}
\mathcal{L}=	\frac{8 c q^2}{K^4}+F-\frac{2^{11/4} K |F|^{5/4}}{5 \sqrt{|q|}}+\mathcal{O}(|F|^{3/2})\,.
\end{multline}

The metric~\eqref{3n23} is solution to the field equations for $x<0$ too but its expansion as $x\to -\infty$ does not have the correct sign in front of $1/x$, thus representing a solution with a negative mass, which is not an interesting case.

For $h=-2$ the polynomial $1+hx+x^2$ has one positive root, $x=1$, at which the NED fields ($F^{tr}$ and $F$), the metric function ($f$) and the Kretschmann scalar ($KS$) diverge. However, the region $-\infty<x<0$ is exempt from any singularity, so that the metric is still given by~\eqref{3n23} and $x$ runs from $-\infty$ (the spatial infinity) to $0$ (the singularity). The expansions of the metric function near the center ($x\to 0^-$) and asymptotically ($x\to -\infty$) are still given by Eqs.~\eqref{exp1} and~\eqref{exp2} on replacing $x$ by $-x$. The ADM mass is still defined by $K=2\pi q^2/(3M)$.\vskip4pt

\section{The case $n=4$: logarithmically singular flat metric\label{secn4}}
\begin{figure}[!htb]
	\centering
	\includegraphics[width=0.99\linewidth]{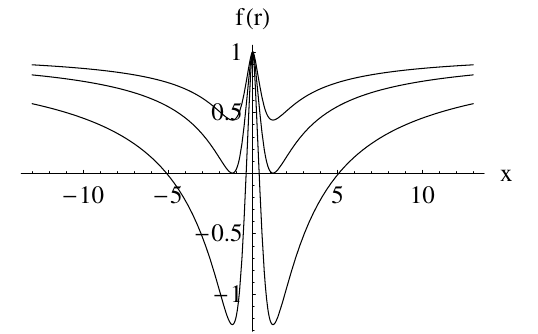}
	\caption{$n=4$ and $h=0$. LS (left-side) and RS (right-side) solutions. Plots of $f(r)$~\eqref{3an4} taking $c=0$ and using the same values of $q^2/M^2$ for the left and right plots. $f(r)$ is regular at the center and assumes a finite value there. Both LS solution and RH solution are asymptotically flat~\eqref{4n4} having the same mass. The plot shows soft LS and RS NSs, extreme LS and RS BHs, and LS and RS BHs each with two horizons. These are the plots of the solution~\eqref{n2h5}-\eqref{n2h6} too corresponding to $n=4$ and $h=2$ and the plots of the fully regular solutions~\eqref{n84a} and~\eqref{n84a2} too corresponding to $n=8,\,h=0$ and $n=8,\,h=2$, respectively.}
	\label{Figlrbhfh02}
\end{figure}
Let
\begin{equation}\label{n4h1}
	\mathcal{L}_F=\frac{1}{x^4}+\frac{h}{x^2}+1\,,
\end{equation}
where we assume that $1+hx^2+x^4=0$ has no positive roots or no roots at all. We obtain
\begin{equation}\label{n4h2}
	F^{tr}=	\frac{qx^2}{K^2(1+hx^2+x^4)}\,,\quad F=\frac{-2q^2x^4}{K^4(1+hx^2+x^4)^2}\,. 
\end{equation}
In this case too we can obtain by integration the general expression of the metric function $f(r)$ in terms of $h$. We have noticed that the Kretschmann scalar has a logarithmic divergence for all $h$ and the expansions of $f(r)$ in the limits of $x\to 0$ and $x\to \infty$ have, up to different coefficients (depending on $h$), the same structure. We will focus on two special cases in the remaining part of this section.

The case $h^2>4$ offers nothing special. Let $h^2<4$, then
\begin{multline}\label{n4h3}
\mathcal{L}=\frac{4 q^2}{K^4} \Big[-\frac{1}{1+h x^2+x^4}+\frac{h}{\sqrt{4-h^2}} \arctan \Big(\frac{h+2 x^2}{\sqrt{4-h^2}}\Big)\\
+\frac{1}{2} \ln \Big(\frac{1+h
	x^2+x^4}{x^4}\Big)+c\Big]\,,\quad h^2<4.
\end{multline}
We see that the case $h=0$ offers a good simplification as well as the case $h=2$, which will be treated separately.\vskip4pt

For $h=0$, Eq.~\eqref{n4h3} reduces to
\begin{equation}\label{n4h4}
	\mathcal{L}=\frac{2 q^2}{K^4} \Big[-\frac{2}{1+x^4}
	+\ln \Big(\frac{1+x^4}{x^4}\Big)+2c\Big]\,,
\end{equation}
along with $F\mathcal{L}_F=-2q^2/[K^4(1+x^4)]$, one can check that,
%\begin{widetext}
\begin{align}
		\label{3an4}	&f(r)=1-\frac{q^2}{K^2}~ p(x)=1-\frac{18M^2}{\pi^2 q^2}~ p(x)\,, \\ 
		\label{3bn4}	&p(x)=\frac{\sqrt{2}}{6 x} \Big[2  [\arctan (1+\sqrt{2} x)-\arctan (1-\sqrt{2} x)]\nonumber\\
		&+ \ln \Big(\frac{1-\sqrt{2}
			x+x^2}{1+\sqrt{2} x+x^2}\Big)+2\sqrt{2}~c x^3 -\sqrt{2}~ x^3 \ln \Big(\frac{x^4}{1+x^4}\Big)\Big]\,,
\end{align}
%\end{widetext}
is a solution to both\footnote{The equality \[\arctan (1+\sqrt{2} x)-\arctan (1-\sqrt{2} x)=\arctan\Big(\frac{\sqrt{2} x}{1-x^2}\Big),\] does not hold for all $x$.} Eqs.~\eqref{3a} and~\eqref{3b}.

%\subsection*{Physical properties}
The polynomials $1\pm\sqrt{2} x+x^2$ in~\eqref{3bn4} do not have real roots. Thus, the metric solution $f(r)$ is regular everywhere except maybe at the center $r=0$. However the presence of the factor $x^3$ in front of the $\ln$ function removes any singularity at the center, as the Maclaurin series justifies it \[f(r)=1-\frac{2q^2 (2+3c-3 \ln  x^2) x^2}{9 K^2}-\frac{q^2 x^6}{7 K^2}+\mathcal{O}(x^8)\,.\] But the Kretschmann scalar diverges logarithmically as
%\begin{widetext}
\begin{multline}\label{7n4}
KS=\frac{16 q^4 [19+18 (c-2) c+18 (1-2c) \ln  x^2+144 \ln ^2 x]}{27 K^8}\\+\frac{16 q^4 (84 c-115-84\ln  x^2) x^4}{63 K^8}+\mathcal{O}(x^8\ln x^2)\,,
\end{multline}
which is softening further compared to the previous cases.
%\end{widetext}
%\begin{figure*}[!htb]
%	\centering
%	\includegraphics[width=0.43\linewidth]{px1.pdf}
%	\includegraphics[width=0.43\linewidth]{fx1.pdf}
%	\caption{Plots of $p(x)$~\eqref{3bn4} and $f(x)$~\eqref{3an4}. The graph of $f(x)$ depicts the three cases (from up to down): horizonless solution, extreme BH with one horizon at $x=1.1973586$, and a BH with two horizons.}
%	\label{Figpx1}
%\end{figure*}

The metric solution is asymptotically flat only if $c=0$ behaving in the limits $r\to\pm\infty$ as
\begin{multline}\label{4n4}
f(r)=1-\frac{2cq^2r^2}{3K^4}-\frac{\sqrt{2} \pi  q^2}{3 K |r|}+\frac{q^2}{r^2}\\-\frac{q^2K^4}{10  r^6}+{\mathcal{O}}\Big(\frac{1}{r^8}\Big)\,,\quad r\to \pm\infty
\end{multline}
It has a de Sitter behavior if $c>0$ and an anti-de Sitter behavior if $c<0$. From the previous equation we obtain $M=\pi q^2/(3\sqrt{2}K)$ or $K=\pi q^2/(3\sqrt{2}M)$.
%\begin{equation}\label{5n4}
%K=\frac{\pi q^2}{3\sqrt{2}M}\simeq 0.74048~\frac{q^2}{M}\,.
%\end{equation}

For $x>0$ the function $p(x)$~\eqref{3bn4} has a maximum value of $0.568363$ at $x=1.19736$ and its graph is similar to that given in Fig~\ref{Fign2hm1}. The extreme BH corresponds to
\begin{equation}\label{ex}
\frac{K^2}{q^2}=\frac{\pi^2 q^2}{18M^2}\simeq 0.568363\to \Big(\frac{q^2}{M^2}\Big)_{\text{crit}} \simeq 1.03657\,.
\end{equation}
For $q^2/M^2>(q^2/M^2)_{\text{crit}}$ we have a horizonless soft NS and for $q^2/M^2<(q^2/M^2)_{\text{crit}}$ the solution has two horizons. Plots of $p(x)$~\eqref{3bn4} and $f(r)$~\eqref{3an4} are similar to those shown in Fig~\ref{Fign2hm1}.

As we see from~\eqref{4n4} both the LS solution and the RS solution are asymptotically flat if $c=0$. The graph of $y=f(x)$ is symmetric with respect to the $y$-axis, as shown in Fig~\ref{Figlrbhfh02}.

The second equation in~\eqref{n4h2}, with $h=0$, can be reversed to express $x^4$ in terms of $F$ then to express $\mathcal{L}$~\eqref{n4h4} in terms of $F$. We have
\begin{equation}\label{r4e}
	x^4=-\frac{K^4F+q^2 +\eta \sqrt{q^4+2K^4q^2F}}{K^4F}\,,\quad \eta=\pm 1\,,	
\end{equation}
yielding
%\[\mathcal{L}_F=\frac{-\epsilon q^2+ \sqrt{q^4+2K^4q^2F}}{K^4F}\,,\]&
\begin{multline}\label{r1}
\mathcal{L}=\frac{2 q^2}{K^4}\bigg[\ln\bigg(-\frac{q^2}{K^4F}+\frac{\eta q^2}{K^4F}\sqrt{1+\frac{2 K^4 F}{q^2}}\bigg)+2c-1\\+\eta \sqrt{1+\frac{2 K^4 F}{q^2}}\bigg]\,.
\end{multline}
For $\eta =+1$, it is  possible to impose the condition $\lim_{F\to 0}\mathcal{L}=F$ upon taking $c=0$; while for $\eta =-1$ it is not possible to satisfy the Maxwellian limit. It is obvious that the limit $\lim_{F\to 0}\mathcal{L}$ depends on $\eta$:
\begin{align}
\label{de3}&\mathcal{L}=\frac{4cq^2}{K^4}+F-\frac{K^4}{4q^2}F^2+\cdots\,,\quad \eta =+1\,,\\
\label{de4}&\mathcal{L}=\frac{2q^2}{K^4}~\Big[2(c-1)+\ln \Big(\frac{2q^2}{K^4}\Big)-\ln|F|\Big]-F\nonumber\\
&\qquad +\frac{K^4}{4q^2}F^2+\cdots\,,\quad \eta =-1\,,
\end{align}
[compare with~\eqref{5pp}-\eqref{5ppp}] where for $\eta=+1$, $x\to\infty$ and for $\eta=-1$, $x\to 0$. In both of these two expansions $F\to 0$, showing that $\mathcal{L}$ is a multi-valued function of $F$ (and a single-valued function of $x$).

Despite these differences in the end-behavior defined by the limit $F\to 0$, the expressions of $\mathcal{L}$~\eqref{n4h4}, for $\eta=\pm 1$, yielded the same metric solution~\eqref{3an4}-\eqref{3bn4} and NED fields~\eqref{n4h2}. This may lead to conclude that different NED Lagragians may lead to the same metric solution and NED fields.

The NED fields are regular for all $-\infty<x<\infty$. \vskip4pt

For $h=2$, the metric function and the NED Lagragian have pretty expressions
\begin{align}
\label{n2h5}&f(r)=1-\frac{q^2}{K^2}p(x)=1-\frac{36M^2}{\pi^2 q^2}p(x)\,,\\
\label{n2h6}&p(x)=\frac{2}{3 x}\Big[(c x^2-1) x+\arctan x + x^3 \ln \Big(\frac{1+x^2}{x^2}\Big)\Big]\,,\nonumber\\
&\mathcal{L}=\frac{4 q^2}{K^4} \Big[-\frac{2+x^2}{(1+x^2)^2}+\ln \Big(\frac{1+x^2}{x^2}\Big)+c\Big]\,.
\end{align}
The NED fields are given by~\eqref{n4h2} on setting $h=2$. For $x>0$ the behaviors at the center and at spatial infinity are, apart from different coefficients, similar to those of the case $h=0$. The Kretschmann scalar has a logarithmic divergence too. The ADM mass is such that $K=\pi q^2/6M$ and the function $p(x)$ has a maximum of $0.326538$ at $x=1.24948$, so that the extreme BH corresponds to $\pi^2 q^2/(36M^2)=0.326538$ with one horizon. For $\pi^2 q^2/(36M^2)>0.326538$ we have a horizonless soft NS and for $\pi^2 q^2/(36M^2)<0.326538$ we have a BH with two horizons. Plots of $p(x)$~\eqref{n2h6} and $f(r)$~\eqref{n2h5} are similar to those shown in Fig~\ref{Fign2hm1}.

In this case too both the LS solution and the RS solution are asymptotically flat if $c=0$. The graph of $y=f(x)$ is symmetric with respect to the $y$-axis and is similar to that shown in Fig~\ref{Figlrbhfh02}.

The NED fields are regular for all $-\infty<x<\infty$.

\section{The case $n=8$: Fully regular solutions\label{secn8}}
Let
\begin{equation}\label{n81}
	\mathcal{L}_F=\frac{1}{x^8}+\frac{h}{x^4}+1\,.
\end{equation}
The NED fields read
\begin{equation}\label{n82}
	F^{tr}=	\frac{qx^6}{K^2(1+hx^4+x^8)}\,,\quad F=-2(F^{tr})^2\,. 
\end{equation}
In this case too we can obtain by integration the general expression of the metric function $f(r)$ in terms of $h$. However, as we shall see and for the purpose of this section, it is enough to focus on the symmetric cases $h=0$ and $h=2$. \vskip4pt

For $h=0$ we have obtained the NED Lagragian, the metric function, and its expansions as given below. The solution has mass $M=\sqrt{\sqrt{2}+1} ~\pi  q^2/(6\times 2^{1/4} K)$.
\begin{equation}\label{n83}
\mathcal{L}(x)=\frac{4 q^2}{K^4} \Big(-\frac{2 x^4+(1+x^8) \arctan  x^4}{2 (1+x^8)}+c\Big)\,,
\end{equation}
\begin{equation}\label{n84a}
f(x)=1-\frac{q^2p(x)}{K^2}=1-\frac{36 \sqrt{2} (\sqrt{2}-1) M^2p(x)}{\pi
	^2 q^2}\,,	
\end{equation}
\begin{align}\label{n84b}
&p(x)=\frac{1}{6 x}\Big\{2 x^3 (2 c-\arctan  x^4)\nonumber\\
&-\sum _{i=0}^3 \cos (7\alpha_i) \ln \Big[1+x^2-2(\cos
\alpha_i) x \Big]\\
&+2 \sum _{i=0}^3 \sin (7\alpha_i) \arctan \Big[(\csc \alpha_i) x-\cot \alpha_i\Big]\Big\}\,,\nonumber\\
&\alpha_i=\frac{(2 i+1) \pi }{8}\,,\nonumber
\end{align}
\begin{equation}\label{n85}
	f(x)=1-\frac{2 c q^2 x^2}{3 K^2}+\frac{q^2 x^6}{7 K^2}+\mathcal{O}(x^7)\,,\quad x\to 0\,,
\end{equation}
\begin{multline}\label{n86}
	f(x)=\frac{(\pi -4 c) q^2 x^2}{6 K^2}+1-\frac{\sqrt{\sqrt{2}+1}~ \pi  q^2}{3\times 2^{1/4} K^2 |x|}\\+\frac{q^2}{K^2 x^2}+\mathcal{O}(x^{-3})\,,\quad x\to
	\pm \infty\,.
\end{multline}
The Kretschmann scalar has no pole of whatsoever at the center and elsewhere
\begin{equation}\label{n87}
	KS=\frac{32 c^2 q^4}{3 K^8}-\frac{64 c q^4 x^4}{3 K^8}+\mathcal{O}(x^6)\,.
\end{equation}
Since the NED fields~\eqref{n82}, and the NED Lagragian itself~\eqref{n83}, are also regular everywhere on the $x$-axis, this solution is fully regular for all $-\infty<x<\infty$.\vskip4pt

The solution is asymptotically flat for $c=\pi/4$. For this value of $c$, the function has two local maxima of equal value $0.468870$ at $x=\pm 1.44815$. For $\pi^2 q^2/[36 \sqrt{2} (\sqrt{2}-1) M^2]>0.468870$, the solution is horizonless metric (not a NS), which may describe a star configuration. For $\pi^2 q^2/[36 \sqrt{2} (\sqrt{2}-1) M^2]<0.468870$, the solution is a BH with two horizons and the extreme case, a BH with double horizon, corresponds to $\pi^2 q^2/[36 \sqrt{2} (\sqrt{2}-1) M^2]=0.468870$. Plots of $f(r)$ have the flat W-shape and are similar to those shown in Fig.~~\ref{Figlrbhfh02}.\vskip4pt

\begin{figure*}[!htb]
	\centering
	\includegraphics[width=0.33\linewidth]{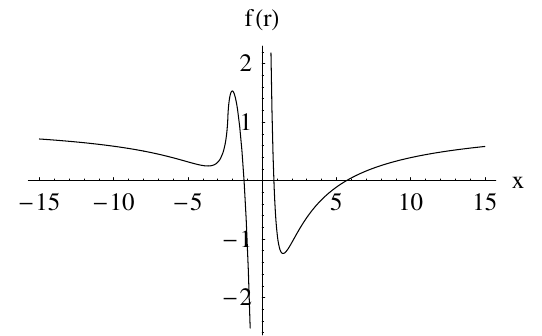}
	\includegraphics[width=0.33\linewidth]{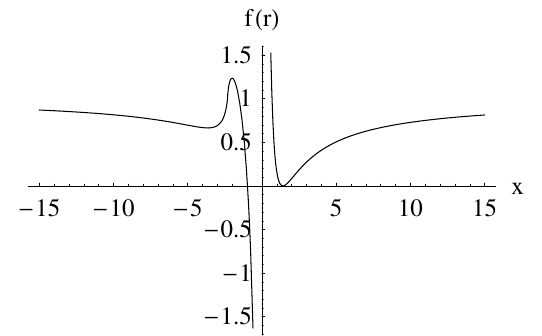}
	\includegraphics[width=0.33\linewidth]{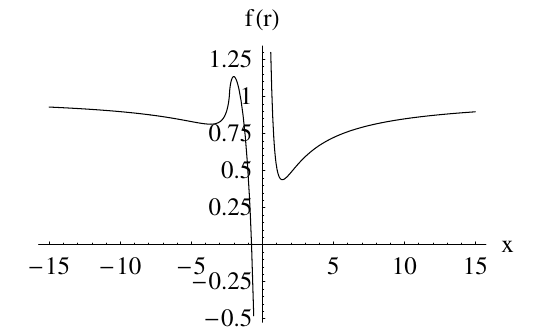}\\
	\includegraphics[width=0.33\linewidth]{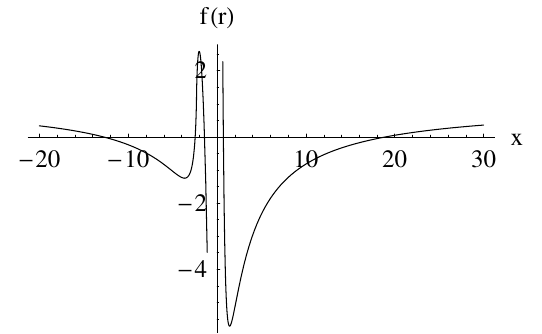}
	\includegraphics[width=0.33\linewidth]{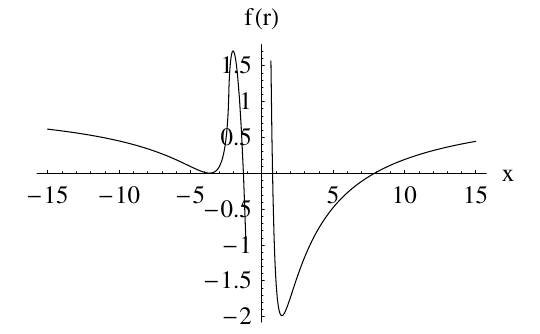}
	\includegraphics[width=0.33\linewidth]{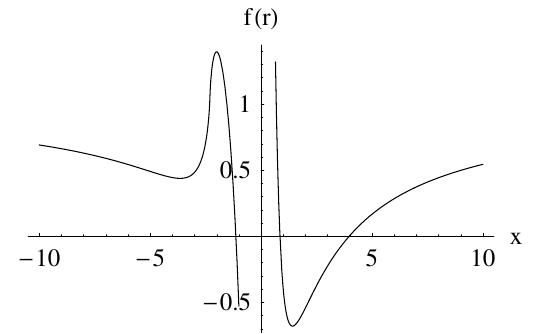}	
	\caption{The case $n=3$. Plots of $f(r)$~\eqref{n350}-\eqref{n35}. Upper Left: LS BH with one horizon and RS BH with two horizons. Upper Middle: LS BH with one horizon and RS BH with one double horizon (extreme case). Upper Right: LS BH with one horizon and RS NS. Lower Left: LS BH with three horizons and RS BH with two horizons. Lower Middle: LS BH with one simple horizon and one double horizon and RS BH with two horizons. Lower Right: Identical to Upper Left plot with a very different value of $K^2/q^2$.}
	\label{Figlrn3}
\end{figure*}

Had we considered a general expression of $\mathcal{L}_F$ [even more general than~\eqref{n81}], we would have obtained non-symmetric solutions as we did in previous sections (compare with Figs.~\ref{Figlrbhfh3} and~\ref{Figlrn3}).\vskip4pt

The behavior of $\mathcal{L}$ was already given in~\eqref{5pp}-\eqref{5ppp}. In full detail we obtain
\begin{align}
\label{de1}&\mathcal{L}=\frac{(4 c-\pi ) q^2}{K^4}+F+\frac{K^8 F^3}{12 q^4}+\cdots\,,\\
\label{de2}&\mathcal{L}=\frac{4 c q^2}{K^4}+3 \Big(\frac{2 q^2}{K^4}\Big)^{2/3}
F^{1/3}-\frac{F}{3}+\cdots\,,
\end{align}
for the behavior at spatial infinity and at the center, respectively. In both of these two expansions $F\to 0$, showing that $\mathcal{L}$ is a multi-valued function of $F$ (and a single-valued function of $x$). A comment on the regularity theorem mentioned in the Introduction is in order: \emph{If $\mathcal{L}$ has two or more expansions in the limit $F\to 0$ and if one of the expansions is Maxwellian, regular metrics (in the sense defined in the theorem) may exist}. In the previous section (and in the next section), $\mathcal{L}$ had (will have) such double expansion~\eqref{de3}-\eqref{de4}, but no regular metric was (will be) found. In this section, we observe a double expansion~\eqref{de1}-\eqref{de2} and a fully regular metric has been obtained~\eqref{n84a}-\eqref{n84b}. The case $h=2$, apart from the shape of the metric and NED fields, is no difference from this case.\vskip4pt

The general metric function $f(r)$ in terms of $|h|< 2$ is given in the Appendix by~\eqref{A00} and the case $h=  2$ is treated separately.\vskip4pt

For $h=2$ the solution has mass $M=\pi q^2/(4\sqrt{2}~ K)$ and
\begin{equation}\label{n832}
\mathcal{L}(x)=\frac{4 q^2}{K^4} \Big(\frac{1- x^4}{2 (1+x^4)^2}+c\Big)\,,
\end{equation}
\begin{equation}\label{n84a2}
f(x)=1-\frac{q^2}{K^2}~p(x)=1-\frac{32 M^2}{\pi
	^2 q^2}~p(x)\,,	
\end{equation}
\begin{align}\label{n84b2}
&p(x)=\frac{\sqrt{2}}{24 x} \Big[6  [\arctan (1+\sqrt{2} x)-\arctan (1-\sqrt{2} x)]\nonumber\\
&+ 3\ln \Big(\frac{1-\sqrt{2}
	x+x^2}{1+\sqrt{2} x+x^2}\Big)+8\sqrt{2}~c x^3 \Big]\,,
\end{align}
{\small
	\begin{equation}\label{n852}
	f(x)=1-\frac{(1+2 c) q^2 x^2}{3 K^2}+\frac{q^2 x^6}{7 K^2}+\mathcal{O}(x^7)\,,\quad x\to 0\,,
	\end{equation}}
\begin{multline}\label{n862}
f(x)=-\frac{2c q^2 x^2}{3 K^2}+1-\frac{\pi  q^2}{2\sqrt{2}~ K^2 |x|}\\+\frac{q^2}{K^2 x^2}+\mathcal{O}(x^{-3})\,,\quad x\to
\pm \infty\,.
\end{multline}
The Kretschmann scalar has no pole of whatsoever at the center and elsewhere
\begin{multline}\label{n872}
KS=\frac{8 (1+2c)^2 q^4}{3 K^8}-\frac{32 (1+2c) q^4 x^4}{3 K^8}\\+\frac{8(229+196c)q^4x^8}{49k^8}+\mathcal{O}(x^{12})\,.
\end{multline}
Since the NED fields ($F^{tr},\,F$), and the NED Lagragian itself~\eqref{n832}, are also regular everywhere on the $x$-axis, this solution is too fully regular for all $-\infty<x<\infty$.

The solution is asymptotically flat for $c=0$. For this value of $c$, the function $p(x)$ has two local maxima of equal value $0.315159$ at $x=\pm 1.67874$. For $\pi^2 q^2/(32 M^2)>0.315159$, the solution is horizonless metric (not a NS), which may describe a star or a soliton. For $\pi^2 q^2/(32 M^2)<0.315159$, the solution is a BH with two horizons and the extreme case, a BH with double horizon, corresponds to $\pi^2 q^2/(32 M^2)=0.315159$. Plots of $f(r)$ have the flat W-shape and are similar to those shown in the right panel of Fig.~\ref{Figlrbhfh02}.

%Had we considered a general expression of $\mathcal{L}_F$ [even more general than~\eqref{n81}], we would have obtained non-symmetric regular solutions as we did in previous sections. 

The two expansions of $\mathcal{L}$ as $F\to 0$~\eqref{5pp}-\eqref{5ppp} are given by
\begin{align}
\label{de12}&\mathcal{L}=\frac{4 c q^2}{K^4}+F-\frac{K^4 F^2}{2 q^2}+\cdots\,,\\
\label{de22}&\mathcal{L}=\frac{2 (1+2c) q^2}{K^4}+3 \Big(\frac{2 q^2}{K^4}\Big)^{2/3}F^{1/3}+ \Big(\frac{2 q^2}{K^4}\Big)^{1/3}F^{2/3}\nonumber\\
&\qquad -\frac{F}{3}+\cdots\,,
\end{align}
where the first expansion is Maxwellian while the solution~\eqref{n84a2} is fully regular.
\begin{figure*}[!htb]
	\centering
	\includegraphics[width=0.43\linewidth]{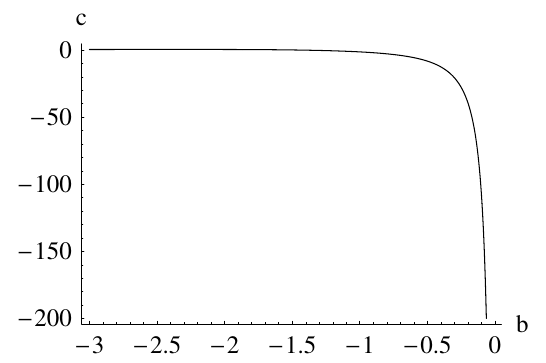}
	\includegraphics[width=0.43\linewidth]{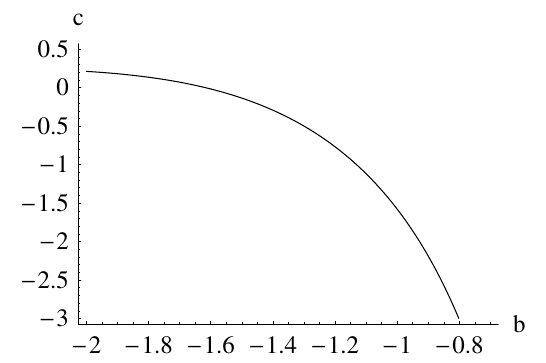}
	\caption{The case $n=3$, $g=a+b$, and $h=(1+a^2b)/a$. Left Panel: A set of possible values of ($b<0\,,c$) and $a=(2+b^3+\sqrt{4-8b^3+b^6})/(2b^2)$ by which both the LS solution and RS solution are asymptotically flat for all $a>0$ and $ab^2<4$. An instance of such a solution is given in Eq.~\eqref{n34}. Right Panel: A zoomed out plot.}
	\label{Figabc}
\end{figure*}

\section{The case $n=3$: Different ADM masses\label{secn3}}
Up to now we have only considered even values of $n$~\eqref{5}. In this section we take $n=3$. This case includes many subcases and yields most solutions we have seen so far and new ones. The general expression of $\mathcal{L}_F$
\begin{equation}\label{n31}
	\mathcal{L}_F=\frac{1}{x^3}+\frac{h}{x^2}+\frac{g}{x}+1\,,
\end{equation}
depends on two real parameters ($h,\,g$). We obtain
\begin{equation}\label{n32}
	F^{tr}=	\frac{qx}{K^2(x^3+gx^2+hx+1)}\,,\; F=-2(F^{tr})^2\,. 
\end{equation}
The subsequent discussion will depend on whether $x^3+gx^2+hx+1=0$ has one simple negative real root or up to three distinct negative roots. In all these cases we have obtained in closed form the expression of the metric function $f(x)$ in terms of $\ln$ and $\arctan$ functions.  

The case with three distinct, two equal, or three equal negative roots of $x^3+gx^2+hx+1=0$ offers nothing special. So, we will focus on the first case ($x^3+gx^2+hx+1=0$ has one simple negative real root), which yields some of the previous solutions seen so far and new ones. In this case $g=a+b$ and $h=(1+a^2b)/a$, where $ab^2<4$ and $a>0$ ($x=-a$ is the unique root of $x^3+gx^2+hx+1=0$). This yields
\begin{equation}\label{n33}
	F^{tr}=	\frac{aqx}{K^2(x+a)(ax^2+abx+1)}\,,\; F=-2(F^{tr})^2\,. 
\end{equation}
The general expression of $\mathcal{L}_F(x)$~\eqref{A1} and $f(x)$~\eqref{A2} are given in the Appendix. Generically, the Kretschmann scalar diverges as $x^{-6}$ at the center and the metric function $f(x)$ diverges as $x^{-1}$. It is not possible to choose the parameters $(a,\,b,\,c)$ so that $f(x)$ is regular at the center and asymptotically flat.

From now on, we consider the special case
\begin{equation}\label{n34}
a=\frac{\sqrt{13}+1}{2}\,,\;  b=-1\,,\; c=\frac{5\sqrt{13}-1}{9}\ln\Big(\frac{\sqrt{13}-1}{6}\Big)\,.
\end{equation}
These values have been fixed on requiring that both the LS solution and RS solution be asymptotically flat for all $a>0$ and $ab^2<4$. These are not the unique values satisfying this requirement. A set of such possible values of ($b<0,\,c$) and $a=(2+b^3+\sqrt{4-8b^3+b^6})/(2b^2)$ is given graphically in Fig.~\ref{Figabc} and it is possibly not the unique set. Given that, the expression of $f(x)$~\eqref{A2} takes the much simple form
\begin{equation}\label{n350}
f(x)=1-\frac{q^2}{K^2}p(x)\,,
\end{equation}
\begin{align}\label{n35}
&p(x)=\frac{2}{9 \Big(7+2 \sqrt{13}\Big) x}\bigg\{2 (2+\sqrt{13}) \sqrt{5+2 \sqrt{13}}~\times\nonumber\\
&\arctan \Big(\frac{\sqrt{5+2 \sqrt{13}}~ (2 x-1)}{3}\Big)\nonumber\\
&-(5+2\sqrt{13}) \ln \Big(\frac{(1+\sqrt{13}+2 x)^2}{4+2 (1+\sqrt{13}) (x-1) x}\Big)\nonumber\\&+6 (7+2 \sqrt{13}) x^2
-\Big[(40+11
	\sqrt{13}) \ln  x^2\nonumber\\
&+\ln \Big(\frac{(1+\sqrt{13})+2 x}{2}\Big)^2-(41+11 \sqrt{13})\times\nonumber\\ 
&\ln \Big(\frac{\sqrt{13}-1+6 (x-1)
		x}{6}\Big)\Big] x^3\bigg\}\,.
\end{align}
This metric diverges at the center as $x^{-1}$ and it is asymptotically flat in both directions ($\pm\infty$) with different masses. Since the asymptotic expansions are sizable, we only provide the ADM masses by
\begin{align}
&M_{(-\infty )}=\frac{q^2}{9 (7+2 \sqrt{13}) K} \Big[(2+\sqrt{13}) \sqrt{5+2 \sqrt{13}}~\pi \nonumber\\
&+2 \sqrt{13} \ln \Big(\frac{\sqrt{13}-1}{6}\Big)-5
\ln \Big(\frac{\sqrt{13}+1}{2}\Big)\Big]\,,\\
&M_{(+\infty )}=\frac{q^2}{9 (7+2 \sqrt{13}) K} \Big[(2+\sqrt{13}) \sqrt{5+2 \sqrt{13}}~\pi \nonumber \\
&-(5+2 \sqrt{13}) \ln
\Big(\frac{\sqrt{13}-1}{6}\Big)\Big]\,,
\end{align}
where $M_{(-\infty )}$ is the mass of the LS solution and $M_{(+\infty )}$ that of the RS solution. Numerically, we obtain $M_{(-\infty )}=0.401508 q^2/K$ and $M_{(+\infty )}=0.560781 q^2/K$.\vskip4pt

The Taylor expansion of $p(x)$ about $x=-a$ includes terms of the form $(x+a)^m\ln(x+a)$ and $m\geq 1$, thus $\lim_{x\to -a^{\pm}}p(x)$ as well as $\lim_{x\to -a^{\pm}}f(x)$ are finite. Numerically, we have $\lim_{x\to -a^{\pm}}p(x)=-0.0130148$. The function $p(x)$~\eqref{n35} has two local maxima of values $0.126466$ and $0.377735$ at the points $x=-3.64633$ and $x=1.41872$, respectively, and a local minimum of value $-0.0887279$ at the point $x=-2.0041$. Since $\lim_{x\to 0^{\pm}} f(x)=\pm \infty$, this means that the LS solution behaves always as a BH with at least one horizon and up to three horizons, while the RS solution may behave as a NS or a BH with one or two horizons. Illustrations are shown in Fig.~\ref{Figlrn3}. As shown in the figure, similar plots may be obtained using very different values of $K^2/q^2$, as is the case with the upper left and lower right plots, which were obtained with $K^2/q^2=0.377735\times 4/9$ and $K^2/q^2=0.126466\times 16/9$, respectively.

The Lagragian $\mathcal{L}$ has double expansion in the limit $F\to 0$ ($\mathcal{L}$ a multi-valued function of $F$) but no regular solution was found [compare with~\eqref{5pp}-\eqref{5ppp}]. Solutions with more than two horizons were also found in~\cite{Odi17,Gao21}.

The NED fields are not regular at $x=-a$.
%\vspace{3mm}

\section{Weak energy condition (WEC)\label{secwek}}
Using the expression of the stress-energy tensor (SET)
\begin{equation}\label{wec1}
	T_\mu^\nu =-2\mathcal{L}_F F_{\mu\sigma}F^{\nu\sigma} + \frac{1}{2}\delta_\mu^\nu \mathcal{L}\,,
\end{equation} 
along with~\eqref{4} we obtain
\begin{equation}\label{wec2}
\rho=-p_r=-F\mathcal{L}_F+\frac{\mathcal{L}}{2}\,,\qquad p_t=-\frac{\mathcal{L}}{2}\,,
\end{equation}
where $\rho$ is the energy density ($=T_t^t$), $p_r$ is the radial pressure ($=-T_r^r$), and $p_t$ is the tangential pressure ($=-T_\theta^\theta=-T_\varphi^\varphi$). The constraints of the WEC, $\rho\geq 0$, $\rho+p_r\geq 0$, and $\rho+p_t\geq 0$, reduce to (compare with~\cite{LB})
\begin{equation}\label{wec3}
	-F\mathcal{L}_F+\frac{\mathcal{L}}{2}\geq 0\quad\text{ and }\quad -F\mathcal{L}_F\geq 0\,.
\end{equation}
[Had we defined the mass function $m(r)$ by $f(r)=1-2m(r)/r$, the two constraints in~\eqref{wec3} would be, by~\eqref{3a}-\eqref{3b}, Eq. (8) and Eq. (9) of~\cite{LB}, respectively].\vskip4pt

Among known regular black holes in NED, there is a set of solutions which violates the WEC and a set which does not (see~\cite{LB} for a detailed list of references). The only regular metrics we have obtained were derived in Sec.~\ref{secn8}, corresponding to $n=8$, and we took $h=0$ and $h=2$. For these values of $h$, $\mathcal{L}_F>0$~\eqref{n81} for all $-\infty<x<\infty$, so that the second condition in~\eqref{wec3} is satisfied, since by~\eqref{4}, $F<0$. This conclusion remains true for all $|h|<2$. 

For ($n=8,\,h=0$) and $c=\pi/4$ [asymptotically flat metric~\eqref{n84a}-\eqref{n84b}], \[\rho=\frac{q^2}{K^4}\Big[\frac{\pi}{2}-\arctan(x^4)\Big]>0,\] ensuring the non-violation of the WEC. This conclusion remains true for all ($n=8,\,|h|<2$) since \[\rho=\frac{q^2}{K^4}~\frac{2}{\sqrt{4-h^2}}\Big[\frac{\pi}{2}-\arctan\Big(\frac{2x^4+h}{\sqrt{4-h^2}}\Big)\Big]>0.\] For this case ($n=8$), the general metric function $f(r)$ in terms of $|h|< 2$ is given in the Appendix by~\eqref{A00}.
%As we shall see most of the solutions derived in this work do not violates the WEC. We will discuss them according the value of $n$. 

For ($n=8,\,h=2$) and $c=0$ [asymptotically flat metric~\eqref{n84a2}-\eqref{n84b2}], \[\rho=\frac{q^2}{K^4}~\frac{1}{x^4+1}>0,\] ensuring the non-violation of the WEC for ($n=8,\,h=2$).

We conclude that all regular metrics satisfying ($n=8,\,-2<h\leq 2$) do not violate the WEC. 

While the solutions for $n=2$ and $n=4$ are not regular, their asymptotically flat counterparts do not violate the WEC if $-2<h\leq 2$.\vskip8pt

\section{Concluding remarks\label{seccon}}
We have obtained spherically symmetric charged solutions endowed with the property that the NED fields are, for some arrangement of parameters, regular for all allowed values of the radial coordinate. The Kretschmann scalar of the associated metric diverges at the center $x=0$ from worst case as $x^{-6}$ to softer case as $\ln x$. Most derived solutions (NSs and/or BHs) have two asymptotic behaviors and the corresponding ADM masses may be different or equal. 

We have shown that multi-valued Lagragians $\mathcal{L}(F)$ may lead to even fully regular solutions besides one of their expansions in the limit $F\to 0$ has Maxwellian limit.

We have obtained five symmetric solutions about the $y$-axis, that is, even functions of $x$. These are given by Eqs.~\eqref{heq0}, \eqref{3an4}, \eqref{n2h5}, \eqref{n84a}, and~\eqref{n84a2}. The first three solutions are naturally matched at the center $x=0$ in such a way that the metric function $f(r)$ and its first derivative are continuous at $x=0$. The last two solutions~\eqref{n84a} and~\eqref{n84a2}, however, are smoothly matched at the center $x=0$ in such a way that the metric function $f(r)$ and all its derivatives are continuous at $x=0$.

For the two fully regular solutions~\eqref{n84a} and~\eqref{n84a2}, particles with zero angular momentum ($L=0$) can go from one spatial infinity to another. If their specific energy (energy per mass) $E\geq 1$, by the radial law of motion \[(\dot{r})^2=E^2-f(r)(1+L^2/r^2),\] where $L$ is the specific angular momentum, which is zero for radial motion, we conclude that $(\dot{r})^2$ remains positive all the way from one spatial infinity to another, since $f(r)\leq 1$ as show in Fig.~\ref{Figlrbhfh02}. Note that the 
%is $(\dot{r})^2=E^2-f(r)$ 
%$(\dot{r})^2=E^2-f(r)(1+L^2/r^2)$ where $L$ is the specific angular momentum, which is zero for radial motion 
effective potential, $f(r)(1+L^2/r^2)$, has the same shape as $f(r)$ if $L=0$ and it can be approximated as a rectangular potential barrier. Particles with $E<1$ can still by quantum tunneling effect cross the center. Knowing that the transition coefficient depends on the potential barrier $V_0$ and on its width $2a$~\cite{book} These parameters are determined as follows: As one can see from Fig~\ref{Figlrbhfh02}, the height of the effective potential, in the vicinity of the center and where $f(r)>0$, is $V_0\sim  1$ and its width $2a$ can be simulated with the distance between the two minima. 
%\vspace{1mm}

%\section{The case $n=5$}
%\begin{equation}\label{1n5}
%	F^{tr}=	\frac{qr^3}{r^5+K^5}=\frac{qx^3}{K^2(x^5+1)}\,,\quad F=-\frac{2q^2x^6}{K^4(x^5+1)^2}\,. 
%\end{equation}
% 
%This yields
%\begin{equation}\label{an1}
%\mathcal{L}=\frac{4 q^2}{K^4} \int \frac{(2 x^5-3)}{(1+x^5)^2} \, \dd x	\,,
%\end{equation}
%\begin{multline}
%\mathcal{L}=\frac{2 q^2}{5 K^4}\Big[-\frac{10 x}{1+x^5}+2 \sqrt{10-2 \sqrt{5}} \arctan \Big(\frac{1+\sqrt{5}-4 x}{\sqrt{10-2 \sqrt{5}}}\Big)\\
%+2 \sqrt{10+2
%	\sqrt{5}} \arctan \Big(\frac{1-\sqrt{5}-4 x}{\sqrt{10+2 \sqrt{5}}}\Big)-\ln [4 (1+x)^4]\\+(1-\sqrt{5}) \ln [2-(1-\sqrt{5})
%x+2 x^2]\\+(1+\sqrt{5}) \ln [2-(1+\sqrt{5}) x+2 x^2]\Big]\,.
%\end{multline}
%With $F\mathcal{L}_F=-2q^2x/[K^4(x^5+1)]$, we have now all necessary tools to check that the metric
%\begin{multline}
%	qaz
%\end{multline}
%is solution to Eqs.~\eqref{3a} and~\eqref{3b}.

%\subsection{Properties of the dynamical normal solution}

%\subsection{Properties of the dynamical phantom solution} 

%\section*{Acknowledgment}

\section*{Appendix: Generic expression of $f(x)$\label{secaa}}
\renewcommand{\theequation}{A.\arabic{equation}}
\setcounter{equation}{0}
In this section we provide the generic expressions of $f(x)$ and other related functions and parameters.\vskip8pt
\subsection*{Case: $n=2$, $|h|< 2$}
Using the expression of $\mathcal{L}(x)$~\eqref{3n2a}, we obtain by integration
{\small
\begin{widetext}
	\begin{align}\label{A0}
		f(x)=1-\frac{2 q^2}{3 K^2} (1-2 h x+c x^2)-\frac{4 q^2 (2-3 h x^3+h^3 x^3)}{3 \sqrt{4-h^2} K^2 x} \arctan \Big(\frac{2 x+h}{\sqrt{4-h^2}}\Big)-\frac{2
			(h^2-1) q^2 x^2}{3 K^2} \ln \Big(\frac{1+x^2+h x}{x^2}\Big)\,.
	\end{align}
\end{widetext}}%
This expression for $h\neq\epsilon 2$ reduces to~\eqref{3n2} if $h=\epsilon$, and reduces to~\eqref{3n22} if $h=\epsilon\sqrt{3}$ with $\epsilon=\pm 1$. For $h=\epsilon 2$, $f(x)$ is given in~\eqref{3n23}.

\subsection*{Case: $n=8$, $|h|< 2$}
Using the expression of $\mathcal{L}(x)$~\eqref{n83}, we obtain by integration $f(x)=1-(q^2/K^2)p(x)$ with
{\small
\begin{equation*}
	p(x)=\frac{2 x^2}{3} \Big[c-\frac{1}{\sqrt{4-h^2}}\arctan \Big(\frac{h+2 x^4}{\sqrt{4-h^2}}\Big)\Big]+\frac{u(x)}{3 x}\,,
\end{equation*}}%
{\small
\begin{widetext}
	\begin{align}\label{A00}
		&u(x)=-\frac{\Big(h-\sqrt{2+h}\Big)}{\sqrt{4-h^2} \sqrt{2-\sqrt{2+h}}} \Big[\arctan \Big(\frac{2 x+\sqrt{2-\sqrt{2+h}}}{\sqrt{2+\sqrt{2+h}}}\Big)+\arctan
		\Big(\frac{2 x-\sqrt{2-\sqrt{2+h}}}{\sqrt{2+\sqrt{2+h}}}\Big)\Big]\nonumber\\
		&-\frac{\sqrt{2+\sqrt{2+h}} \Big(4+h-3 \sqrt{2+h}\Big)}{\sqrt{4-h^2} \Big(2-\sqrt{2+h}\Big)}
		\Big[\arctan \Big(\frac{2 x+\sqrt{2+\sqrt{2+h}}}{\sqrt{2-\sqrt{2+h}}}\Big)+\arctan \Big(\frac{2 x-\sqrt{2+\sqrt{2+h}}}{\sqrt{2-\sqrt{2+h}}}\Big)\Big]\\
		&+\frac{\sqrt{2+\sqrt{2+h}}
			\Big(4+h-3 \sqrt{2+h}\Big)}{2 \sqrt{4-h^2} \Big(2-\sqrt{2+h}\Big)} \ln \Big(\frac{1+\sqrt{2-\sqrt{2+h}} x+x^2}{1-\sqrt{2-\sqrt{2+h}} x+x^2}\Big)+\frac{\Big(h-\sqrt{2+h}\Big)}{2
			\sqrt{4-h^2} \sqrt{2-\sqrt{2+h}}} \ln \Big(\frac{1+\sqrt{2+\sqrt{2+h}} x+x^2}{1-\sqrt{2+\sqrt{2+h}} x+x^2}\Big)\nonumber\,.
	\end{align}
\end{widetext}}%
The generic expression of the ADM mass is
\[ M=\frac{\Big(\sqrt{2+h}-\sqrt{2-h}+\sqrt{4-h^2}-h\Big) \pi  q^2}{6 \sqrt{\Big(4-h^2\Big) \Big(2-\sqrt{2+h}\Big)} K}. \] Taking $h=0$, we recover the expression of $f(x)$~\eqref{n84a}-\eqref{n84b} and $M$ given in Sec~\ref{secn8}.

\subsection*{Case: $n=3$, $g=a+b$, $h=(1+a^2b)/a$}
%\vspace{-25.3mm}
{\small
\begin{widetext}
\begin{align}\label{A1}
\mathcal{L}(x)=&\frac{4 q^2}{K^4}\bigg[\frac{1}{x}+\frac{1}{(1+a^3-a^2 b) (a+x)}+\frac{a^2 (1+a (a-b) (b+x))}{(1+a^3-a^2 b) [1+a x (b+x)]}-\frac{2
	a^2 \sqrt{a} [3 b+a^2 b^2-a (2+b^3)]}{(1+a^3-a^2 b) \sqrt{4-a b^2}} \arctan \Big(\frac{\sqrt{a} (b+2 x)}{\sqrt{4-a
		b^2}}\Big)\nonumber\\
\quad &+\frac{(1+a^2 b) \ln  x^2}{a}-\frac{\ln (a+x)^2}{a (1+a^3-a^2 b)}-\frac{a^2 (1+a^2 b-a b^2) \ln [1+ax (b+x)]}{1+a^3-a^2 b}+c\bigg]\,.
\end{align}
\end{widetext}
\begin{widetext}
\begin{align}\label{A2}
f(x)=&1-\frac{q^2}{K^2}p(x)\,,\nonumber\\
p(x)=&\frac{4 x}{3}-\frac{2}{3 (1+a^3-a^2 b) x} \bigg[2 (a^2 b-2) \sqrt{\frac{a}{4-a b^2}} \arctan
\Big(\sqrt{\frac{a}{4-a b^2}} (b+2 x)\Big)+a^2 \ln \Big(\frac{(a+x)^2}{1+a x (b+x)}\Big)\bigg]\nonumber\\
\quad &-\frac{2 x^2}{3 (1+a^3-a^2 b)}\bigg[c(a^2 b-a^3-1)+2a^2 \sqrt{\frac{a}{4-a b^2}}~[3 b-2a+(a-b) ab^2] \arctan \Big[\sqrt{\frac{a}{4-a b^2}} (b+2
x)\Big]\nonumber\\
\quad &-\frac{1+a^3 [1+a (a-b) b]}{a} \ln  x^2+\frac{1}{a} \ln  (a+x)^2+a^2 [1+a (a-b) b] \ln [1+a x (b+x)]\bigg]\,.
\end{align}
\end{widetext}}

%
%\section*{Appendix: Nonvanishing expressions of $F^{\mu\nu}$\label{secaa}}
%\renewcommand{\theequation}{A.\arabic{equation}}
%\setcounter{equation}{0}

%\section*{Appendix B: ??? $\pmb{???}$\label{secab}}
%\renewcommand{\theequation}{B.\arabic{equation}}
%\setcounter{equation}{0}

%\newpage

\end{document}